\documentclass[aps,twocolumn,showpacs,preprintnumbers,amsmath,amssymb,superscriptaddress,floatfix,nofootinbib]{revtex4}

\usepackage{graphicx}
\usepackage{epsfig}
\usepackage{epstopdf}
\usepackage{hyperref}
\usepackage{amsmath}
\usepackage{amsfonts}
\usepackage{amssymb}

\usepackage{graphicx,color,dcolumn,booktabs}
\usepackage{txfonts}
\usepackage{amssymb}
\usepackage{indentfirst}
\usepackage{subfig}
\usepackage{float}
\usepackage{tabularx} 

\allowdisplaybreaks
\begin{document}

\title{The $\rho (\omega) B^* (B)$ interaction and states of $J=0,1,2$.}

\author{P.~ Fern\'andez Soler}
\affiliation{Departamento de
F\'{\i}sica Te\'orica and IFIC, Centro Mixto Universidad de
Valencia-CSIC Institutos de Investigaci\'on de Paterna, Aptdo.
22085, 46071 Valencia, Spain}

\author{Zhi-Feng Sun}
\affiliation{Departamento de
F\'{\i}sica Te\'orica and IFIC, Centro Mixto Universidad de
Valencia-CSIC Institutos de Investigaci\'on de Paterna, Aptdo.
22085, 46071 Valencia, Spain}

\author{J. Nieves}
\affiliation{IFIC, Centro Mixto Universidad de
Valencia-CSIC Institutos de Investigaci\'on de Paterna, Aptdo.
22085, 46071 Valencia, Spain}

\author{E.~Oset}
\affiliation{Departamento de
F\'{\i}sica Te\'orica and IFIC, Centro Mixto Universidad de
Valencia-CSIC Institutos de Investigaci\'on de Paterna, Aptdo.
22085, 46071 Valencia, Spain}

\date{\today}

\begin{abstract}
In this work, we study systems composed of a $\rho/\omega$ and $B^*$ meson pair. We find three bound states in isospin, spin-parity channels $(1/2, 0^+)$, $(1/2, 1^+)$ and $(1/2, 2^+)$. The state with $J=2$ can be a good candidate for the $B_2^*(5747)$. We also study the $\rho B$ system, and a bound state with mass $5728$ MeV and width around $20$ MeV is obtained, which can be identified with the $B_1(5721)$ resonance. In the case of $I=3/2$, one obtains repulsion and thus, no exotic (molecular) mesons in this sector are generated in the approach.
\end{abstract}

\maketitle

\section{Introduction}

Chiral symmetry, reflecting the QCD dynamics at low energies, has played a crucial role in the description of the hadron interactions. Originally developed for the interaction of pseudoscalar mesons \cite{Gasser:1983yg} and of the meson nucleon system \cite{Ecker:1994gg,Bernard:1995dp}, the need to incorporate vector mesons into the framework gave rise to the local hidden gauge approach \cite{hidden1,hidden2,hidden4}, which incorporates the information of the chiral Lagrangians of  \cite{Gasser:1983yg} and extends it to accommodate the vector interaction with pseudoscalars and with themselves. Another important step to understand the dynamics of hadrons at low and intermediate energies was given by incorporating elements of nonperturbative physics, restoring two body unitarity in coupled channels, which gave rise to the chiral unitary approach, that has been instrumental in explaining many properties of hadronic resonances, mesonic \cite{npa,ramonet,kaiser,markushin,juanito,rios} and baryonic \cite{Kaiser:1995cy,angels,ollerulf,carmenjuan,hyodo,ikeda,cola,Borasoy:2005ie,Oller:2005ig,
Oller:2006jw,Borasoy:2006sr,Hyodo:2008xr,Roca:2008kr}.  Concerning the interaction of vector mesons in this unitary approach, the first work was done in \cite{raquelvec}, where surprisingly the $f_2(1270), ~f_0(1370)$ resonances appeared as a consequence of the interaction of $\rho$ mesons from the solution of the Bethe Salpeter equation with the potential generated from the local hidden gauge Lagrangians \cite{hidden1,hidden2,hidden4}.
The generalization to SU(3) of that work was done in \cite {gengvec} and further resonances came from this approach, the $f'_2(1525),f_0(1710)$ among others. Most of these findings were confirmed in the SU(6) spin-flavor symmetry scheme followed by \cite{GarciaRecio:2010ki} and \cite{Garcia-Recio:2013uva}. The step to incorporate charm in the local hidden gauge approach of Refs. \cite{raquelvec,gengvec} was given in \cite{hidekoraquel}, and the interaction of $\rho$, $\omega$ and $D^*$ was studied extrapolating to the charm sector the local hidden gauge approach. Three $D$ states with spin $J=0,1,2$ were obtained, the second one identified with the $D^*(2640)$ and the last one with the $D^*_2(2460)$. The first state, with $J=0$, was predicted at $2600$ MeV with a width of about 100 MeV. This state is also in agreement with the $D(2600)$, with a similar width, reported after the theoretical work in \cite{delAmoSanchez:2010vq}.  The properties of these resonances are well described by the theoretical approach.

The success in the predictions of this theoretical framework in the light and the charm sectors suggests to give the step to the bottom sector and make predictions at this early stage. The extension is straightforward, because the interaction in the local hidden gauge approach is provided by the exchange of vector mesons. The exchange of light vectors is identical to the case of the $\rho D^*$ interaction, since the $c$ or $b$ quarks act as spectators. In the exchange of heavy vectors, the form and the coefficients are also the same, since  the $\bar B$  meson can be obtained from the
$D$  simply replacing the $c$ quark by the $b$ quark. However, instead of exchanging a $D^*$ in the subdominant terms, one exchanges now a $B^*$ meson. These terms are anyway subdominant. Hence, it is not surprising that the predictions that we obtain in this work in the $B$ sector are very similar to those obtained in \cite{hidekoraquel} in the $D$ sector.

We shall also discuss the heavy quark spin symmetry (HQSS), which we show is satisfied by the dominant terms of the interaction, and then discuss the behaviour of the HQSS breaking terms subdominant HQSS breaking terms.
 We make predictions for three states from the  $\rho /\omega\, B^*$ interaction and compare with available experimental states.

In a similar way, we also deal with the interaction of $\rho B$ in s-wave, which gives rise to a state of $J=1$ which we can identify with a state already existing. This interaction follows also from the local hidden gauge approach, although equivalent chiral Lagrangians have been used in the light sector \cite{Birse:1996hd,roca,GarciaRecio:2010ki,Garcia-Recio:2013uva} and in the $D$ sector \cite{daniaxial,christoph}, where also QCD sum rules have been investigated \cite{Torres:2013saa}.

\section{Formalism}
We are going to use the local hidden gauge approach where the interaction is given mainly by the exchange of vector mesons. We follow closely the approach of \cite{hidekoraquel} and with mi\-ni\-mal changes we can obtain most of the equations.

\subsection{Vector-vector interaction}
We take the vector-vector interaction from \cite{hidden1} as
\begin{equation}
\mathcal{L}_{III}=-\frac{1}{4}\left\langle V_{\mu\nu}V^{\mu\nu}\right\rangle.\label{lagrangian}
\end{equation}
where the symbol $\langle\rangle$ represents the trace in SU(4) flavor space (we consider $u, d, s$ and $b$ quarks), with
\begin{equation}
V_{\mu\nu}=\partial_\mu V_\nu-\partial_\nu V_\mu-ig[V_\mu,V_\nu]
\end{equation}
and
\begin{equation}
V_{\mu}=\left(
             \begin{array}{cccc}
               \frac{\rho^0}{\sqrt{2}}+\frac{\omega}{\sqrt{2}} & \rho^+ & K^{*+} & B^{*+} \\
               \rho^- & -\frac{\rho^0}{\sqrt{2}}+\frac{\omega}{\sqrt{2}} & K^{*0} & B^{*0} \\
               K^{*-} & \bar{K}^{*0} & \phi & B_s^{*0} \\
               B^{*-} & \bar{B}^{*0} & \bar{B}_s^{*0} & \Upsilon \\
             \end{array}\label{matrix}
           \right)_{\mu}
\end{equation}
standing for the vector representation of the different $q\bar{q}$ pairs, and the coupling $g$ is given by
\begin{equation}
g=\frac{M_V}{2f}
\end{equation}
with the pion decay constant $f\simeq 93$ MeV, and $M_V\simeq 770$ MeV.

\begin{figure}[htb]
	\centering
	\subfloat[]{
	\includegraphics[scale=0.7]{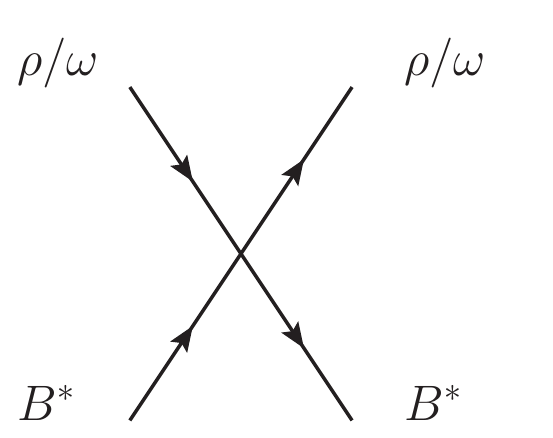}}
	\subfloat[]{
	\includegraphics[scale=0.7]{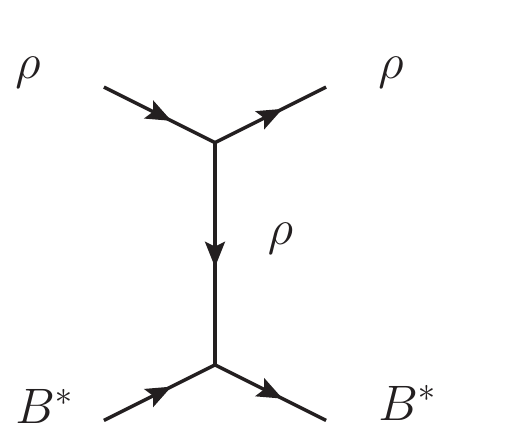}}\\
	\subfloat[]{
	\includegraphics[scale=0.7]{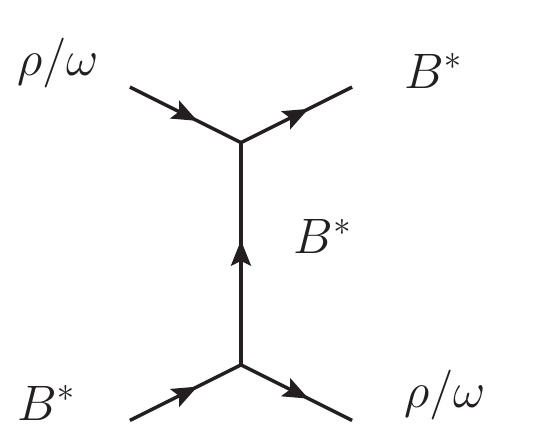}}
\caption{The model for the $\rho/\omega\, B^*$ interaction.}
\label{contact}	
\end{figure}

The local hidden gauge Lagrangians also contains a four vector contact term
\begin{equation}
\mathcal{L}_{III}^{(c)}=\frac{g^2}{2}\langle V_\mu V_\nu V^{\mu}V^{\nu}-V_\nu V_\mu V^{\mu}V^{\nu}\rangle,
\end{equation}
which in the $\rho B^*$ channel gives rise to the term depicted in Fig. \ref{contact}(a).
\\
From Eq. (\ref{lagrangian}) we also get a three vector interaction term
\begin{equation}
\mathcal{L}_{III}^{3V}=ig\left\langle \left(\partial_\mu V_\nu-\partial_\nu V_\mu\right)V^\mu V^\nu\right\rangle.
\end{equation}
This latter Lagrangian gives rise to a $VV$ interaction term through the exchange of a virtual vector meson, as depicted in Figs. \ref{contact}(b) and \ref{contact}(c).
As in \cite{hidekoraquel} we also assume that the three momenta of the particles are small compared to the vector masses. This helps to simplify the formalism.

We consider the $\rho/\omega\,B^*$ states
\begin{align}
\left|\rho B^*; I=1/2, I_3=1/2\right\rangle&=-\sqrt{\frac{2}{3}}\left|\rho^+B^{*0}\right\rangle-\sqrt{\frac{1}{3}}\left|\rho^0B^{*+}\right\rangle, \nonumber\\
\left|\rho B^*; I=3/2, I_3=3/2\right\rangle&=-\left|\rho^+ B^{*+}\right\rangle,\nonumber\\
\left|\omega B^*; I=1/2, I_3=1/2\right\rangle&=\left|\omega B^{*+}\right\rangle.
\end{align}
where we use the phase convention where the isospin doublets are $(K^{*+},K^{*0})$, $(\bar{K}^{*0},-K^{*-})$, $(B^{*+},B^{*0})$, $(\bar{B}^{*0},-B^{*-})$ and the rho triplet is $(-\rho^+, \rho^0, \rho^-)$.

The contact terms are all of the type
\begin{align}
-it^{(c)}_{\rho B^*\to \rho B^*}&=-ig^2(\alpha\epsilon_\mu^{(1)}\epsilon_\nu^{(2)}\epsilon^{(3)\nu}\epsilon^{(4)\mu}\nonumber\\
&+\beta\epsilon_\mu^{(1)}\epsilon_\mu^{(2)}\epsilon^{(3)\nu}\epsilon^{(4)\nu}\nonumber\\
&+\gamma\epsilon_\nu^{(1)}\epsilon_\mu^{(2)}\epsilon^{(3)\nu}\epsilon^{(4)\mu}),
\end{align}
where $\epsilon^\mu$ are the polarization vectors of the vector mesons in the order 1, 2, 3, 4, where these indices are used in the reaction  $1+2\to 3+4$.

Analoguosly, the terms associated to vector exchange of the type of Fig. \ref{contact}(b) are particularly easy, since, neglecting the external three momenta, these terms are of the type
\begin{align}
t^{(ex)}&=\frac{g^2}{M_V^2}\alpha^\prime(k_1+k_3)\cdot (k_2+k_4)\epsilon_\mu\epsilon_\nu\epsilon^\mu\epsilon^\nu.
\label{Eq:exchange-term}
\end{align}
Then we can separate these terms into the different contribution of spin (we work only with angular momentum $L=0$) which are given by \cite{raquelvec}
\begin{align}
\mathcal{P}(0)&=\frac{1}{3}\epsilon_\mu \epsilon^\mu \epsilon_\nu \epsilon^\nu,\\
\mathcal{P}(1)&=\frac{1}{2}(\epsilon_\mu \epsilon_\nu \epsilon^\mu \epsilon^\nu-\epsilon_\mu \epsilon_\nu \epsilon^\nu \epsilon^\mu),\\
\mathcal{P}(2)&=\frac{1}{2}(\epsilon_\mu \epsilon_\nu \epsilon^\mu \epsilon^\nu+\epsilon_\mu \epsilon_\nu \epsilon^\nu \epsilon^\mu)-\frac{1}{3}\epsilon_\mu \epsilon^\mu \epsilon_\nu \epsilon^\nu.
\end{align}
We can see that, while the contact terms give rise to different combinations of spin, the vector exchange term of type of Fig. \ref{contact}(b), contains the sum $\mathcal{P}(0)+\mathcal{P}(1)+\mathcal{P}(2)$, with equal weights for the different spins. This combination, corresponding to the exchange of a light vector meson ($\rho$, $\omega$, $\phi$, $K^*$) if allowed, sa\-tis\-fies HQSS to which we shall come back later on. On the other hand, the exchange of a heavy vector meson contains the combination
\begin{equation}
\epsilon_\mu \epsilon_\nu \epsilon^\nu \epsilon^\mu=\mathcal{P}(0)-\mathcal{P}(1)+\mathcal{P}(2)
\end{equation}
and does not satisfy leading order HQSS constraints, as we shall see in Sect. \ref{subsection:Results-rho-B*-system}. This goes in line with HQSS, since the exchange of heavy vectors is penalized versus the exchange of light ones by a factor $M_V^2/M_{B^*}^2$ from the pro\-pa\-ga\-tors and become subdominant. The contact term is also subdominant since it goes like $M_V/M_{B^*}$ of the dominant term from the exchange of a light vector. HQSS is satisfied only for the dominant term in the $\mathcal{O}(\frac{1}{M_{B^*}})$ counting, as it should be.

\subsection{Vector-pseudoscalar interaction}
\begin{figure}[htb]
\begin{tabular}{cc}
\includegraphics[scale=0.7]{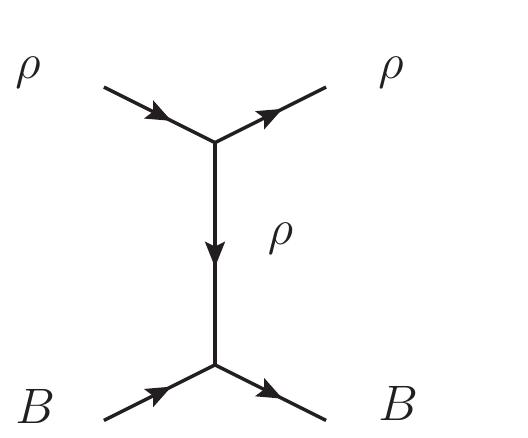}
\end{tabular}
\caption{Interaction of $\rho B$ with vector exchange.\label{exchange3}}
\end{figure}
We shall also consider the $\rho B$ interaction. This proceeds via the exchange of a vector meson as in Fig. \ref{exchange3} and in this case there is no contact term.
One can see that in the limit (which we also take) that $q^2/M_V^2\to 0$, where $q$ is the momentum transfer, one obtains the chiral Lagrangian of \cite{Birse:1996hd}.
The lower vertex $VBB$ is given by the Lagrangian provided by the extended local hidden gauge approach
\begin{align}
\mathcal{L}=-ig \left\langle \left[\phi, \partial_\mu \phi \right]V^\mu\right\rangle,\label{lagrangian2}
\end{align}
where now $\phi$ is the corresponding matrix of Eq. (\ref{matrix}) for $q\bar{q}$ in the pseudoscalar representation. We obtain the same ex\-pre\-ssion as for $\rho B^*\to \rho B^*$ (direct term in Fig. \ref{contact}(b)) replacing
\begin{align}
\epsilon_\mu \epsilon_\nu \epsilon^\mu \epsilon^\nu\to -\epsilon_\mu \epsilon^\mu.
\end{align}
Then up to the factor $-\epsilon_\mu \epsilon^\mu\to \vec{\epsilon}\cdot\vec{\epsilon}^{\,\prime}$, a scalar factor, that becomes unit in the only possible spin state here which is $J=1$ with $L=0$, we find the same potential for $\rho B\to \rho B$ as for $\rho B^*\to \rho B^*$ with the dominant light vector exchange in any spin channel.

\section{Decay modes of the $\rho B^*$ channels}

\begin{figure}[htb]
\begin{tabular}{cc}
{\scalebox{0.7}{\includegraphics{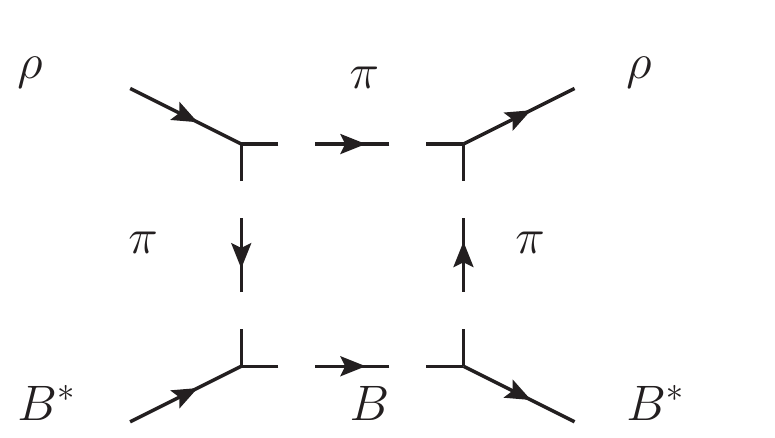}}}
\end{tabular}
\caption{Box diagram to account for the decay of $\rho B^*$ into $\pi B$ state.\label{box}}
\end{figure}
As in \cite{hidekoraquel} we take into account the box diagrams of the type of Fig. \ref{box}. Note that these diagrams do not exist for the case of the $\rho B\to \rho B$ interaction because we would need a $\pi BB$ vertex which does not exist. The details are identical as those in \cite{hidekoraquel} (section VI) by simply changing the masses of the particles $D^*$, $D$ by those of the $B^*$, $B$ mesons.

From the time of \cite{hidekoraquel} one has learned something relevant concerning the $B^* B \pi$ vertex.
As discussed in \cite{Liang:2014eba}, the $B^*B\pi$ and $K^*K\pi$ are identical at the quark level but the use of the $(2\omega)^{(1/2)}$ normalization of the meson fields require that the e\-ffec\-ti\-ve vertex of Eq. (\ref{lagrangian2}) is renormalized to become
\begin{equation}
g_{B^*B\pi}=g\frac{m_{B^*}}{m_{K^*}}.\label{eq:g}
\end{equation}
Then we use this vertex in the box diagram instead of the empirical one $g^\prime_{D^*D\pi}$ used in \cite{hidekoraquel}.

\section{Heavy quark spin symmetry considerations\label{section:HQSS-considerations}}
Let us consider the $\rho B^{(*)}$ meson pair. In the particle basis we have four states for each isospin combination, namely $\left|\rho B,\,J=0\right\rangle$,  $\left|\rho B^*,\,J=0\right\rangle$,  $\left|\rho B^*,\,J=1\right\rangle$ and  $\left|\rho B^*,\,J=2\right\rangle$. In the HQSS basis \cite{Xiao:2013yca}, the states are classified in terms of the quantum numbers: $J$, total spin of the meson pair system and $\cal L$, total spin of the light quark system. In addition, for this particular simple case in the HQSS basis, the total spin of the heavy quark subsystem, $S_Q$, is fixed to $1/2$, as well as the spin of the light quarks and heavy quarks in each of the two mesons. Thus, the four orthogonal states in the HQSS basis are given by $\left|\mathcal{L}=1/2,\,J=0\right\rangle$, $\left|\mathcal{L}=1/2,\,J=1\right\rangle$, $\left|\mathcal{L}=3/2,\,J=1\right\rangle$ and $\left|\mathcal{L}=3/2,\,J=2\right\rangle$. In all the cases the spin of the $\bar{b}$-antiquark, $S_Q$, is coupled to $\cal L$ to give $J$. The a\-ppro\-xi\-ma\-te HQSS of QCD leads at leading order (LO), i.e., neglecting $\mathcal{O}\left(\Lambda_{\text{QCD}}/m_Q\right)$ to important simplifications when the HQSS basis is used. 
\begin{align}
 \left\langle \mathcal{L}',J';\alpha'\right| H^{\text{QCD}}\left| \mathcal{L},J;\alpha\right\rangle=\delta_{\alpha \alpha'}\delta_{\cal{L}\cal{L}'}\delta_{J J'}\mu^{\alpha}_{2\cal{L}}
\end{align}
where $\alpha$ stands for other quantum numbers (isospin and hypercharge), which are conserved by QCD. The reduced matrix elements, $\mu^{\alpha}_{2\cal{L}}$, depend only on the spin (parity) of the light quark subsystem, $\cal{L}$, and on the additional quantum numbers, $\alpha$, that for the sake of simplicity we will omit in what follows.\\
The particle and HQSS bases are easily related through 9-j symbols (see  \cite{Xiao:2013yca}), and one finds
\begin{align}
 \begin{split}
 \left|\rho\,B,\,J=1	\right\rangle = -\sqrt{\frac{1}{3}}\left| \mathcal{L}=1/2,\,J=1 \right\rangle+\sqrt{\frac{2}{3}}\left| \mathcal{L}=3/2,\,J=1\right\rangle,\\
 \left|\rho\,B^*,\, J=0	\right\rangle = \left| \mathcal{L}=1/2,\,J=2 \right\rangle, \\
 \left|\rho\,B^*,\,J=1	\right\rangle =\sqrt{\frac{2}{3}}\left| \mathcal{L}=1/2,\,J=1 \right\rangle+\sqrt{\frac{1}{3}}\left| \mathcal{L}=3/2,\,J=1\right\rangle, \\
 \left|\rho\,B^*,\, J=2	\right\rangle = \left| \mathcal{L}=3/2,\,J=2 \right\rangle.  
 \end{split}
\end{align}
we obtain, in the infinite heavy quark mass limit,
\begin{align}
&\left\langle \rho B\right\vert H^{\text{QCD}} \left\vert \rho B\right\rangle = \frac{1}{3}\mu_1+\frac{2}{3}\mu_3,\\
&\left\langle \rho B^*,J=0\right\vert H^{\text{QCD}} \left\vert \rho B^*,J=0\right\rangle = \mu_1,\\
&\left\langle \rho B^*,J=1\right\vert H^{\text{QCD}} \left\vert \rho B^*,J=1\right\rangle = \frac{2}{3}\mu_1+\frac{1}{3}\mu_3,\\
&\left\langle \rho B^*,J=2\right\vert H^{\text{QCD}} \left\vert \rho B^*,J=2\right\rangle = \mu_3,\\
&\langle \rho B\vert H^{\text{QCD}} \vert \rho B^*,J=1\rangle = -\frac{\sqrt{2}}{3}\mu_1+\frac{\sqrt{2}}{3}\mu_3 .
\end{align}
Since we have not coupled the $\rho B$ with $\rho B^*$ in our model because it involves anomalous terms which are very small in this case, then $\mu_1 =\mu_3$ and we conclude that all the matrix elements are equal for $\rho B^*$ in $J =0,1,2$ and also for $\rho B$.
We can see that the dominant term for the light vector exchange (Eq. \ref{Eq:exchange-term}) fulfils the rules of HQSS relations, but the contact term and $B^*$ exchange, which are subdominant in the $\left(\frac{1}{m_{B^*}}\right)$ counting, do not satisfy those relations, since they do not have to (note that when rewriting this potential in the usual normalization of HQSS, we would have an extra $\left(\frac{1}{2\omega_{B^*}}\right)$ factor that makes the $\rho$ exchange to go like $\left(\frac{1}{m_{B^*}}\right)^0$, the contact term like $\left(\frac{1}{m_{B^*}}\right)$ and the $B^*$ exchange like $\left(\frac{1}{m_{B^*}}\right)$).

\section{Results}
\subsection{Bethe Salpeter resummation}
As in \cite{hidekoraquel}, we resum the diagrams of the Bethe Salpeter series to obtain the scattering matrix $T$ in coupled channels by using
\begin{equation}
T=[1-VG]^{-1}V,
\label{Eq:Bethe-Salpeter}
\end{equation}
where V is the potential $\rho B^*\to \rho B^*$, $\rho B^*\to \omega B^*$, $\omega B^*\to \omega B^*$ that one obtains using the former sections, and $G$ is the vector-vector loop function used in this type of studies and also given explicitly in \cite{hidekoraquel}. All the relevant matrix elements can be obtained from tables I, II and III of \cite{hidekoraquel}. The finite width of the $\rho$ meson is also explicitly taken into account by considering the $\rho$ mass distribution in the construction of the $G$ function.

In the next section we shall discuss our results for both the $\rho B^*$ and $\rho B$ systems by using the coupled channel unitary approach, where we only consider the contribution of s-wave. The interaction in the $I=3/2$ case is repulsive, and thus in what follows we will focus in the $I=1/2$ sector. 

\begin{figure*}[htb]
\centering
\includegraphics[scale=0.35]{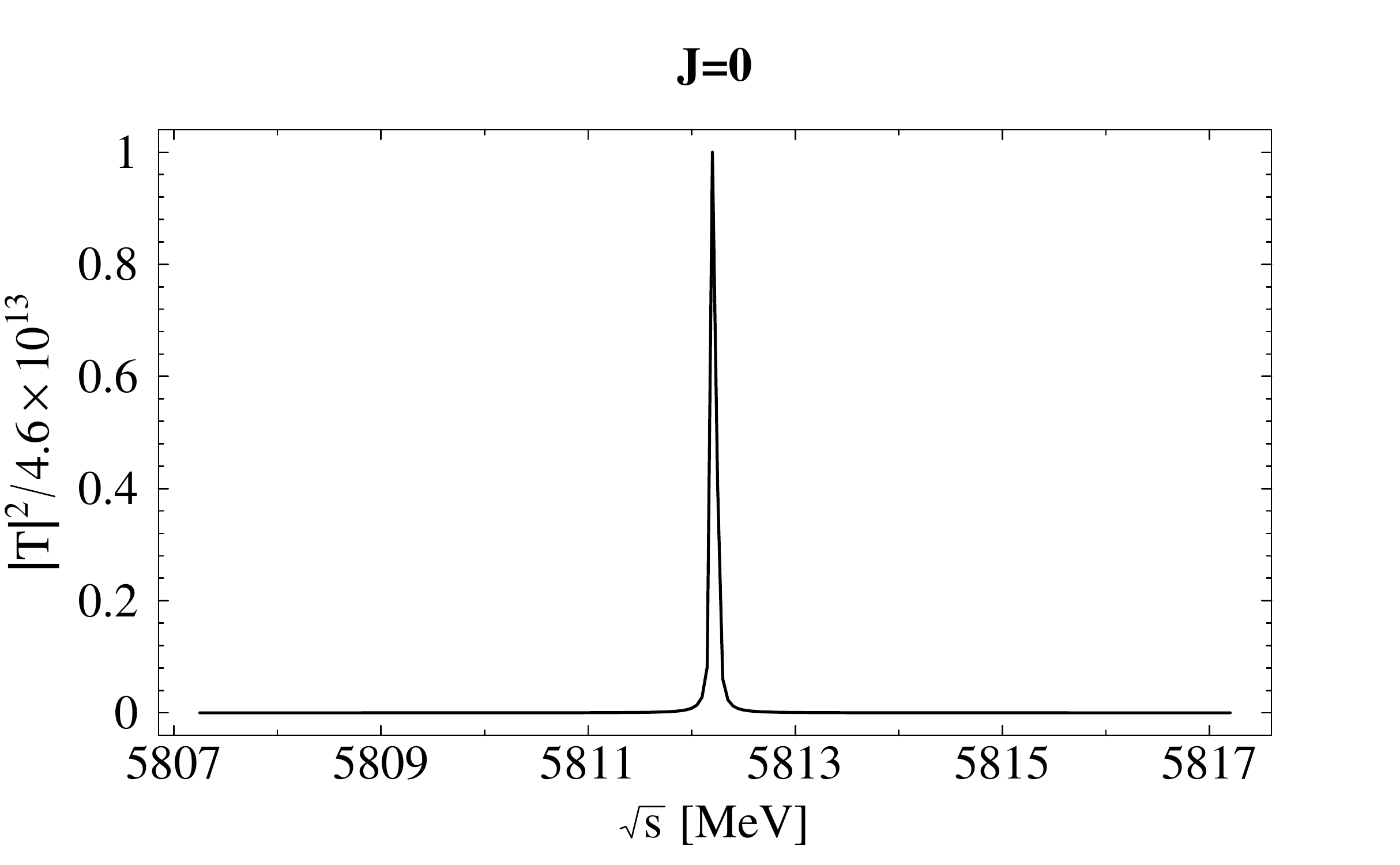} 
\includegraphics[scale=0.35]{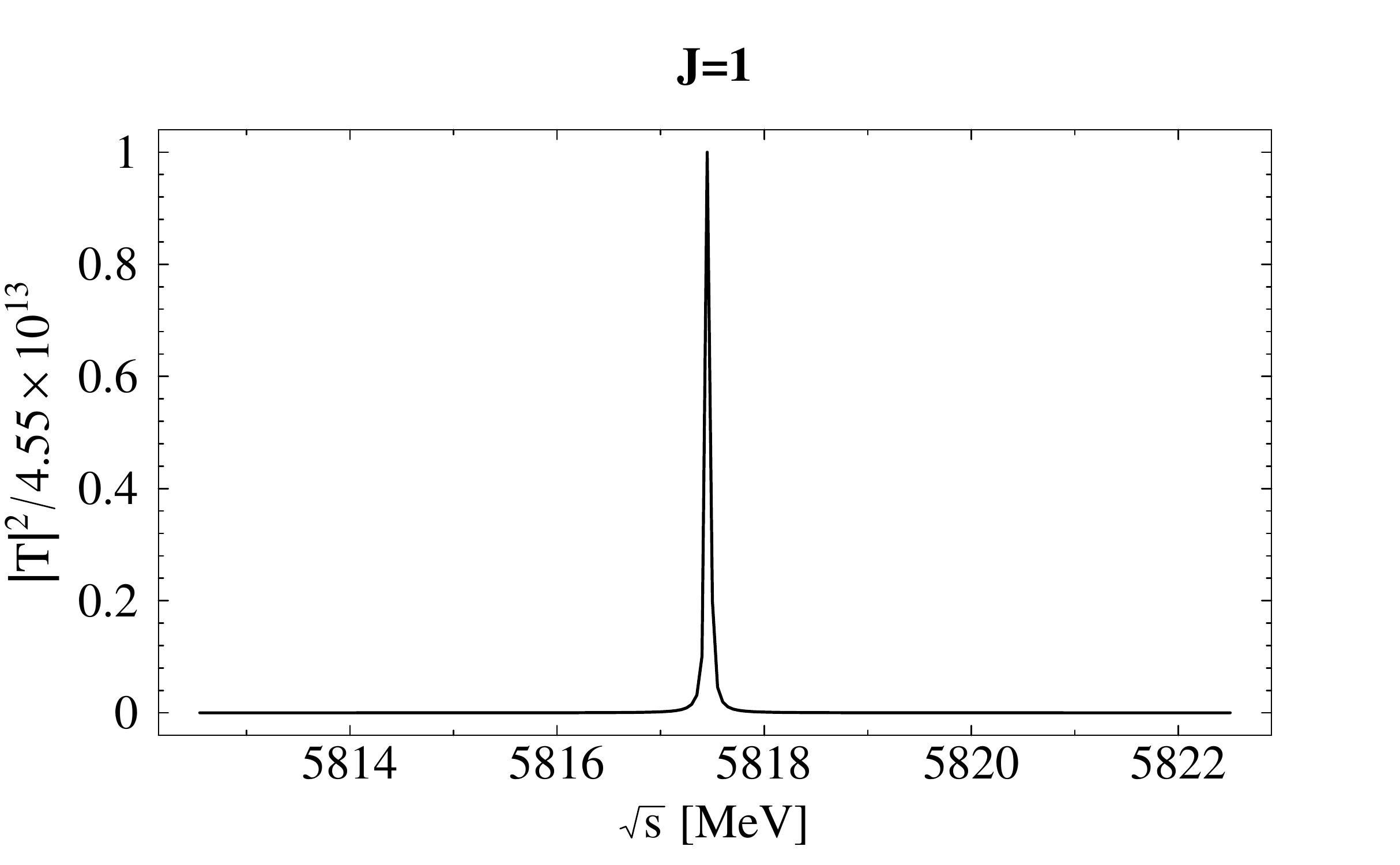}
\includegraphics[scale=0.35]{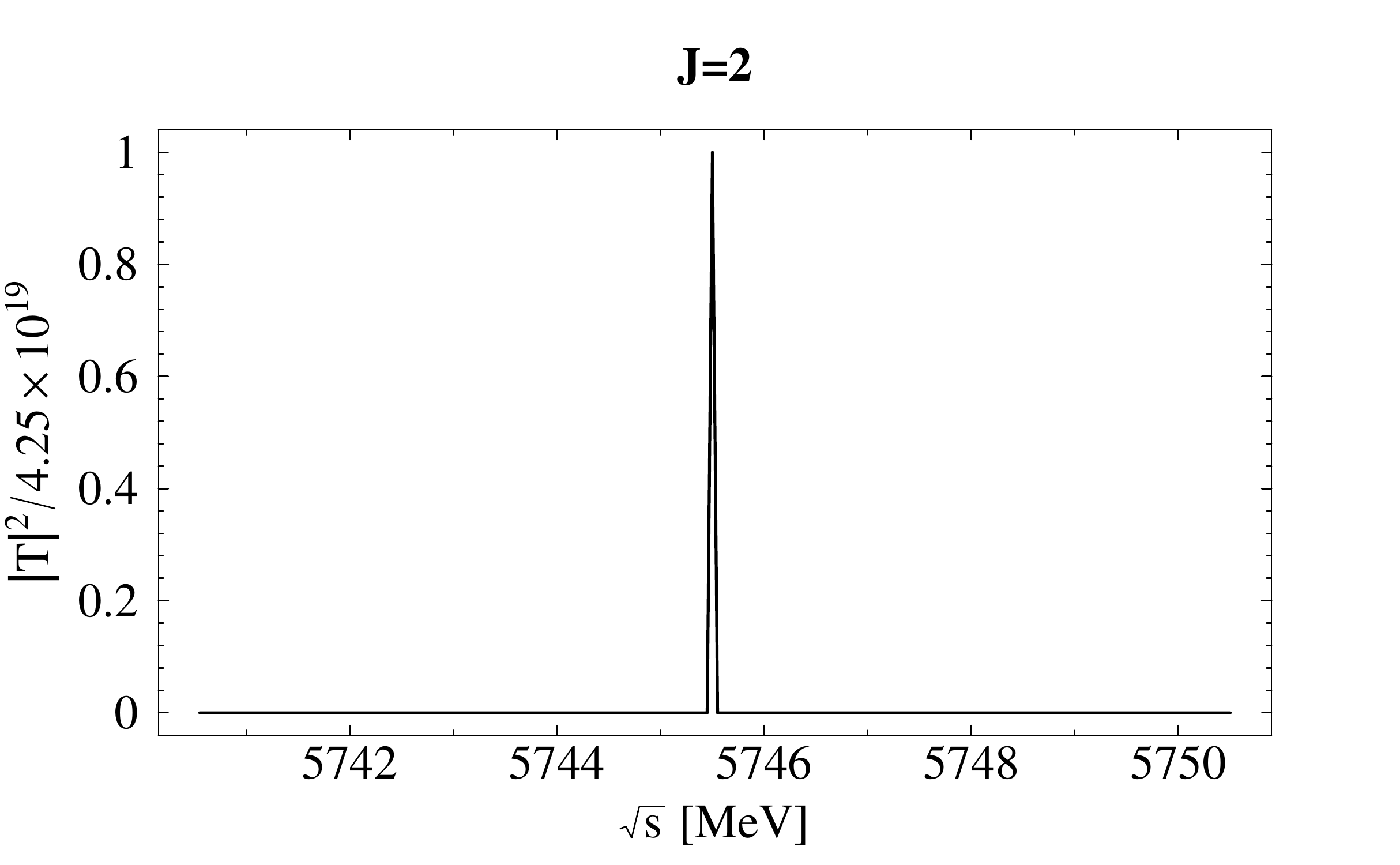}
\caption{Squared amplitude for $I=1/2$ and $J=0,1,2$ including the convolution of $\rho$ mass distribution.\label{TsquareRhoConvolution}}
\end{figure*}

\subsection{$\rho/\omega\, B^*$ system \label{subsection:Results-rho-B*-system}}
In the first step, we introduce the kernel or potential $V$, co\-rres\-pon\-ding to the contact and vector exchange contributions. We can get an intuitive idea of the results by using the results of Table I of \cite{hidekoraquel}, adapted to the present case in Table \ref{Table:Potential-rhoBx-rhoBx}.
\begin{table}[t]
\caption{$V(\rho B^*\to \rho B^*)$ with isospin $1/2$ for the different spin channels including the exchange of one heavy vector meson. Here $\kappa=m_\rho^2/m_{B^*}^2$.}
\begin{tabular}{ccccc}
\toprule[1pt]
$I$   & $J$ & contact & $\rho$ exchange & $B^*$ exchange \\\hline
1/2 & 0 & 5$g^2$& $-2\frac{g^2}{M_{\rho}^2}(k_1+k_3)\cdot (k_2+k_4)$&$-\frac{1}{2}\frac{\kappa g^2}{M_{\rho}^2}(k_1+k_4)\cdot (k_2+k_3)$\\
1/2 & 1 & $\frac{9}{2}g^2$&$-2\frac{g^2}{M_{\rho}^2}(k_1+k_3)\cdot (k_2+k_4)$&$+\frac{1}{2}\frac{\kappa g^2}{M_{\rho}^2}(k_1+k_4)\cdot (k_2+k_3)$\\
1/2 & 2 & $-\frac{5}{2}g^2$&$-2\frac{g^2}{M_{\rho}^2}(k_1+k_3)\cdot (k_2+k_4)$&$-\frac{1}{2}\frac{\kappa g^2}{M_{\rho}^2}(k_1+k_4)\cdot (k_2+k_3)$\\
\bottomrule[1pt]
\label{Table:Potential-rhoBx-rhoBx}
\end{tabular}
\end{table}

By calculating the potential at the threshold of $\rho B^*$, summing the contact, $\rho$ exchange and $B^*$ exchange contributions we get potentials with weights ($\kappa$ of \cite{hidekoraquel} is now $M_\rho^2/M_{B^*}^2$) $-51g^2$, $-50g^2$, $-58g^2$ for $J=0, 1, 2$, respectively. These results correspond to $-16g^2$, $-14.5g^2$, $-23.5g^2$ of \cite{hidekoraquel}. The strength is bigger than for the $\rho D^*$ system because of the bigger masses of the heavy quarks and we still find that the strength is bigger for $J=2$. However, we also see that the weight for different spins are now more similar in accordance with HQSS as discussed in Sect. \ref{section:HQSS-considerations}.

With the potentials evaluated as a function of the energy as given in Tables I, II, III of \cite{hidekoraquel} we solve the Bethe Sapeter equation (\ref{Eq:Bethe-Salpeter}) in the $\rho B^*$, $\omega B^*$ coupled channels though the contribution of the $\omega$ channel is fairly small. We need to re\-gu\-la\-ri\-ze the G function and use the cut off prescription using $q_{max}=1.3$ GeV. The G function is also convoluted with the $\rho$ mass distribution as in \cite{hidekoraquel}. With this prescription we obtain three bound states for $J=0, 1, 2$ that we plot in Fig. \ref{TsquareRhoConvolution}. The value of $q_{max}$ has been chosen to obtain a mass of $5745$ MeV within the range of $5743\pm 5$ MeV of the nominal mass of the $B_2^*(5747)$ state \cite{Agashe:2014kda}. The masses for the other two states are then predictions: we obtain a state with $J=0$ at $5812$ MeV and another one for $J=1$ at $5817$ MeV.
Here, we can see that the mass of the spin 1 state is larger than that of spin 2, while in the PDG, the resonance $B_1(5721)$ with spin 1 has a mass smaller than the mass of $B_2^*(5747)$. Henceforth, the state with spin 1 that we obtain presents some difficulties to be identified with the $B_1(5721)$. One possibility is that it could be the resonance generated by $\rho B$ interaction, which we shall discuss later. Note that the LO HQSS relation $\mu_1=\mu_3$ has some $1/m_Q$ correction.

The T matrix element close to a pole behaves like
\begin{equation}
T_{ij}\approx \frac{g_ig_j}{z-z_R}
\end{equation}
where $g_i$ is the coupling to channel $i$ ($i=\rho B^*, \omega B^*$) and $z$, $z_R$ are the complex values for $s$ and the resonance position $s_R$. We can get the coupling of one channel as
\begin{align}
g_i^2=\lim_{z\to z_R}T_{ii}(z-z_R)\label{gi2}.
\end{align}
We choose the $\rho B^*$ coupling with positive sign, and for the other channels we use 
\begin{align}
\frac{g_i}{g_j}&=\displaystyle{\lim_{z\to z_R} \frac{T_{ii}}{T_{ij}}},
\end{align}
\begin{table}[b]
\centering
\caption{Couplings of the bound states to the $\rho B^*$ and $\omega B^*$ channels with $I=1/2$ and $J=0,1,2$ in units of GeV. The imaginary parts of the couplings are negligible, less than 0.05 \% of the real part in all the cases.\label{coupling}}
\begin{tabular}{ccccccccccccc}
\toprule[1pt]
channel      & $J=0$&$J=1$&$J=2$\\\midrule[0.5pt]
$\rho B^*$   & $39.6$&$39.3$&$43.6$\\
$\omega B^*$ &$1.0$&$-2.1$&$-2.4$\\
\bottomrule[1pt]
\end{tabular}
\end{table}
which gives us the relative sign for the $\omega B^*$ channel.
Note here that the right hand side of Eq. (\ref{gi2}) is the residue of the amplitude $T_{ii}$.
The coupling to the different channels are listed in Table \ref{coupling}.

The $\rho$ mass distribution is also involved via the convoluted $G$ function and should give a width different from zero to the states. Nonetheless, we obtain that the widths for $J=0$, $1$ and $J=2$ are much smaller than one MeV (see Fig. \ref{TsquareRhoConvolution}). Ho\-we\-ver, in the PDG the width of the $B_2^*(5747)$ state is $23^{+5}_{-11}$ MeV, which is larger than the one obtained here for the state with spin 2. To reconcile the difference, the $\pi B$ decay channel must be included.

The energies of the resonances are close to the threshold of $\rho$ and $B^*$ and far away from that of $\pi$ and $B$. We do not need to treat the $\pi B$ as a coupled channel, since it does not have much weight compared to the $\rho B^*$ and $\omega B^*$ channels. Henceforth, as in \cite{hidekoraquel}, one can compute the box diagrams that are mediated by $\pi B$ and put them in the potential $V$ in order to get the width.
The $\rho B^*$ contribution corresponding to the box diagram was shown in Fig. \ref{box}. We use directly the result of Eq. (41) of \cite{hidekoraquel} and have

\begin{align}
V^{\pi B}&=g^4(\epsilon_i^{(1)}\epsilon_i^{(2)}\epsilon_j^{(3)}\epsilon_j^{(4)}
+\epsilon_i^{(1)}\epsilon_j^{(2)}\epsilon_i^{(3)}\epsilon_j^{(4)}
+\epsilon_i^{(1)}\epsilon_j^{(2)}\epsilon_j^{(3)}\epsilon_i^{(4)})\nonumber\\
&\times \frac{8}{15\pi^2}\int_0^{q_{max}}dq\vec{q}^{\,\,6}\left(\frac{1}{2\omega_\pi}\right)^3\left(\frac{1}{k_1^0+2\omega_\pi}\right)^2\nonumber\\
&\times\frac{1}{k_2^0-\omega_\pi-\omega_B+i\epsilon}\frac{1}{k_4^0-\omega_\pi-\omega_B+i\epsilon}\frac{1}{k_1^0-2\omega+i\epsilon}\nonumber\\
&\times\frac{1}{k^0_3-2\omega_\pi+i\epsilon}\frac{1}{P^0-\omega_\pi-\omega_B+i\epsilon}\frac{1}{P^0+\omega_\pi+\omega_B}\nonumber\\
&\times\left(\frac{1}{k_2^0+\omega_\pi+\omega_B}\right)^2\frac{1}{2\omega_B}f(P^0,\vec{q}^{\,2})\label{boxintegral}
\end{align}
\begin{figure*}
  \centering
  \includegraphics[scale=0.35]{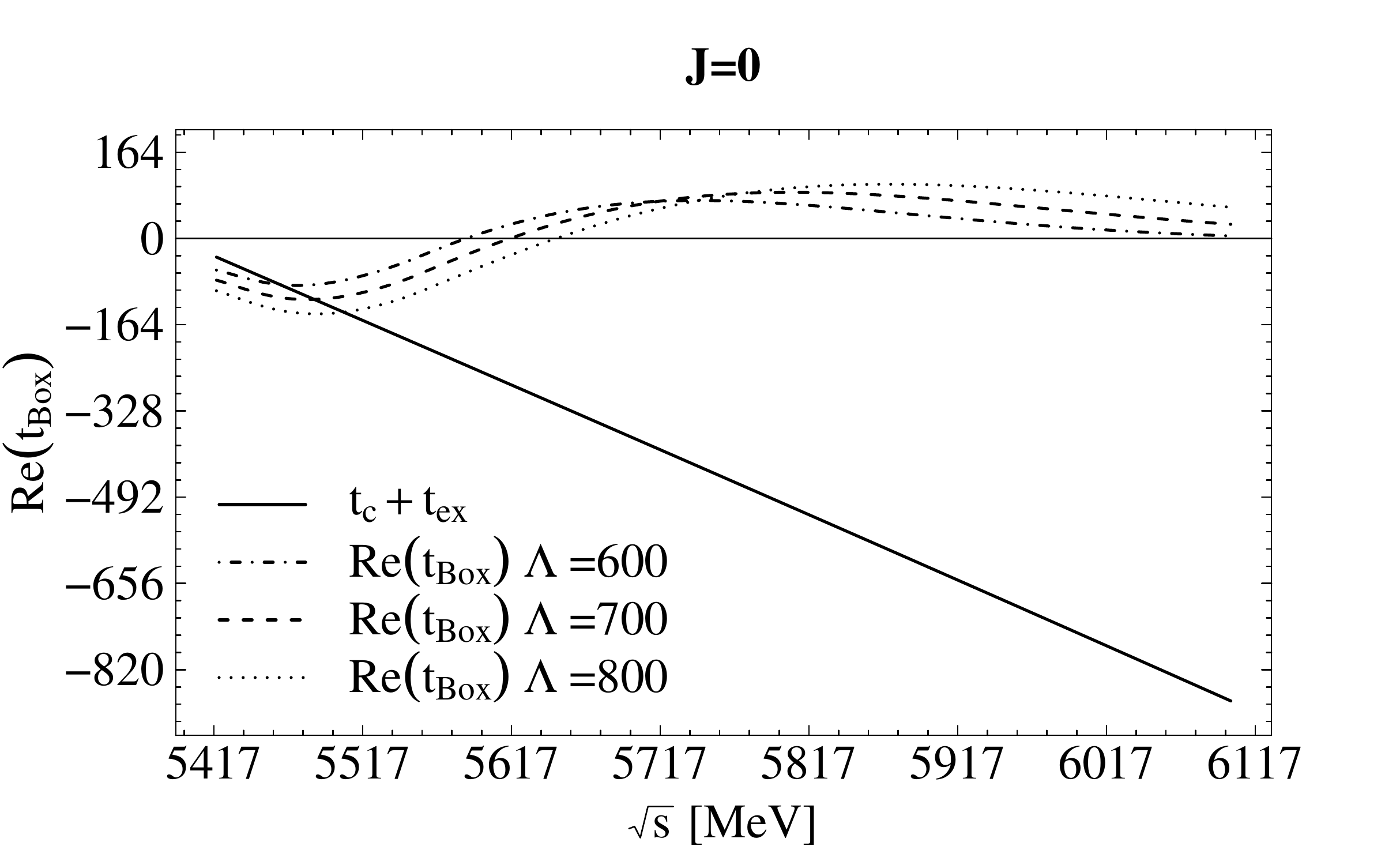}
  \includegraphics[scale=0.35]{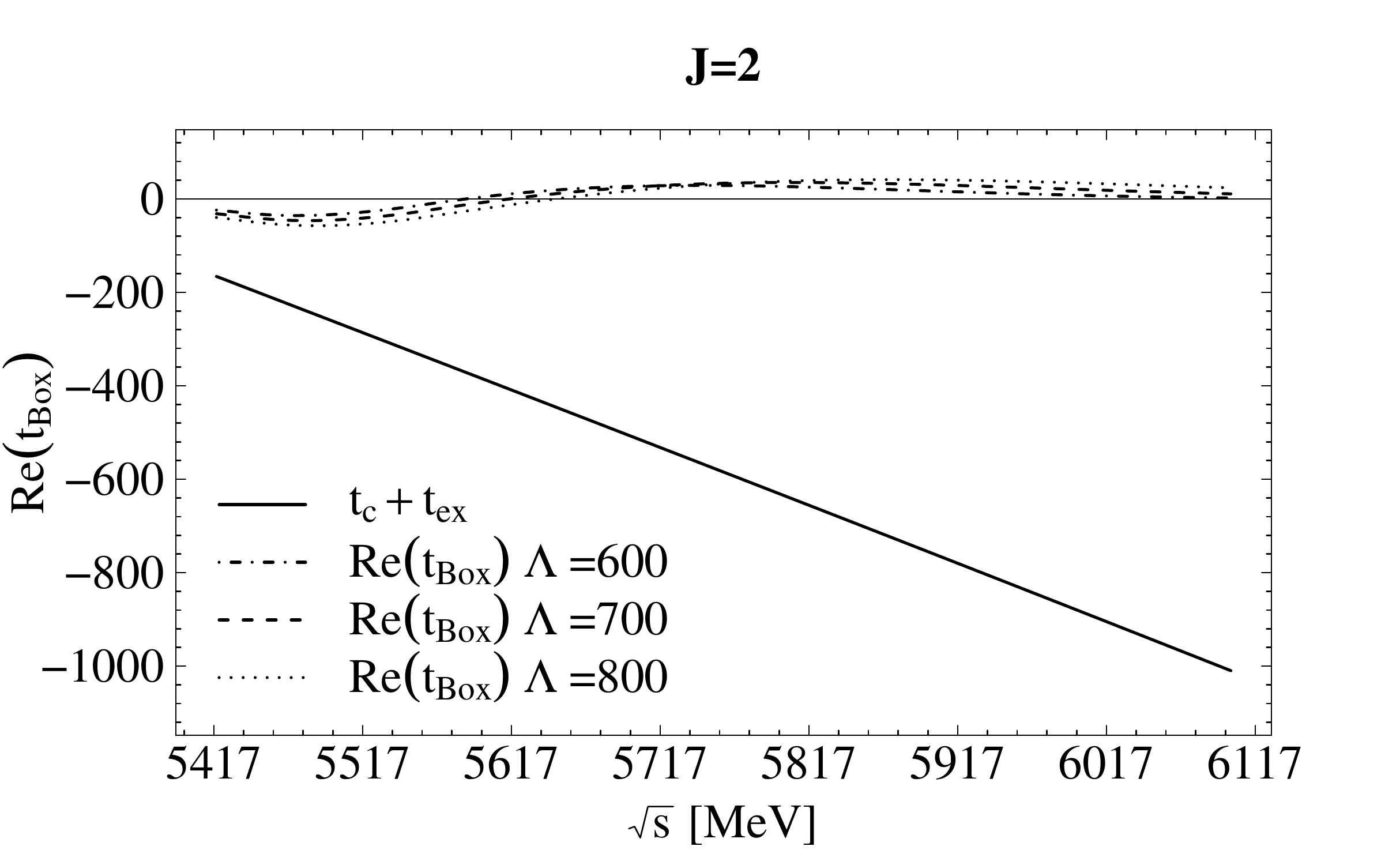}
  \caption{The real part of box potential for $(I,J)=(1/2,0)$ and $(I,J)=(1/2,2)$ compared with those from contact and vector exchange terms.\label{realpartbox}}
\end{figure*}

\begin{figure*}
  \centering
  \includegraphics[scale=0.35]{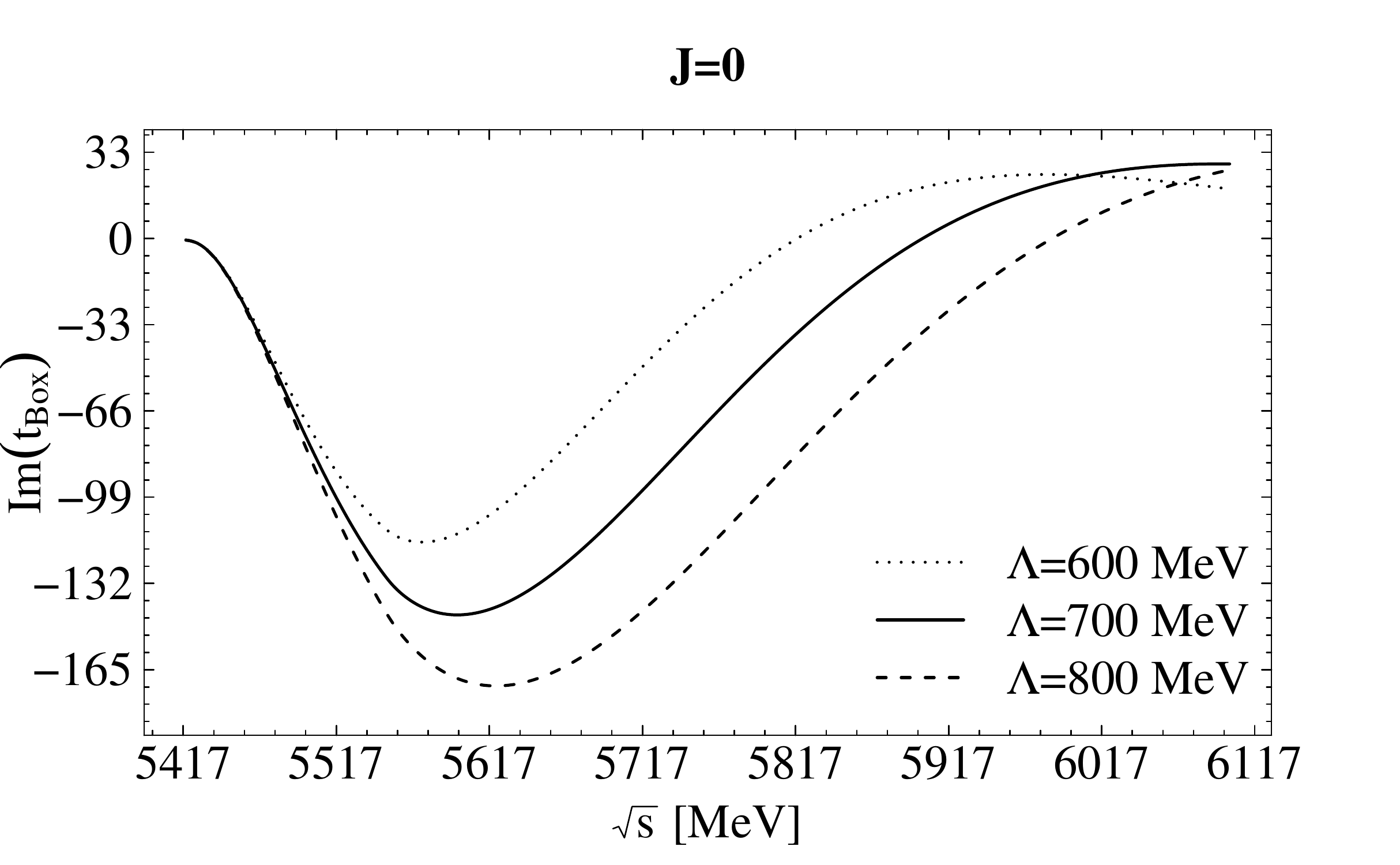}
  \includegraphics[scale=0.35]{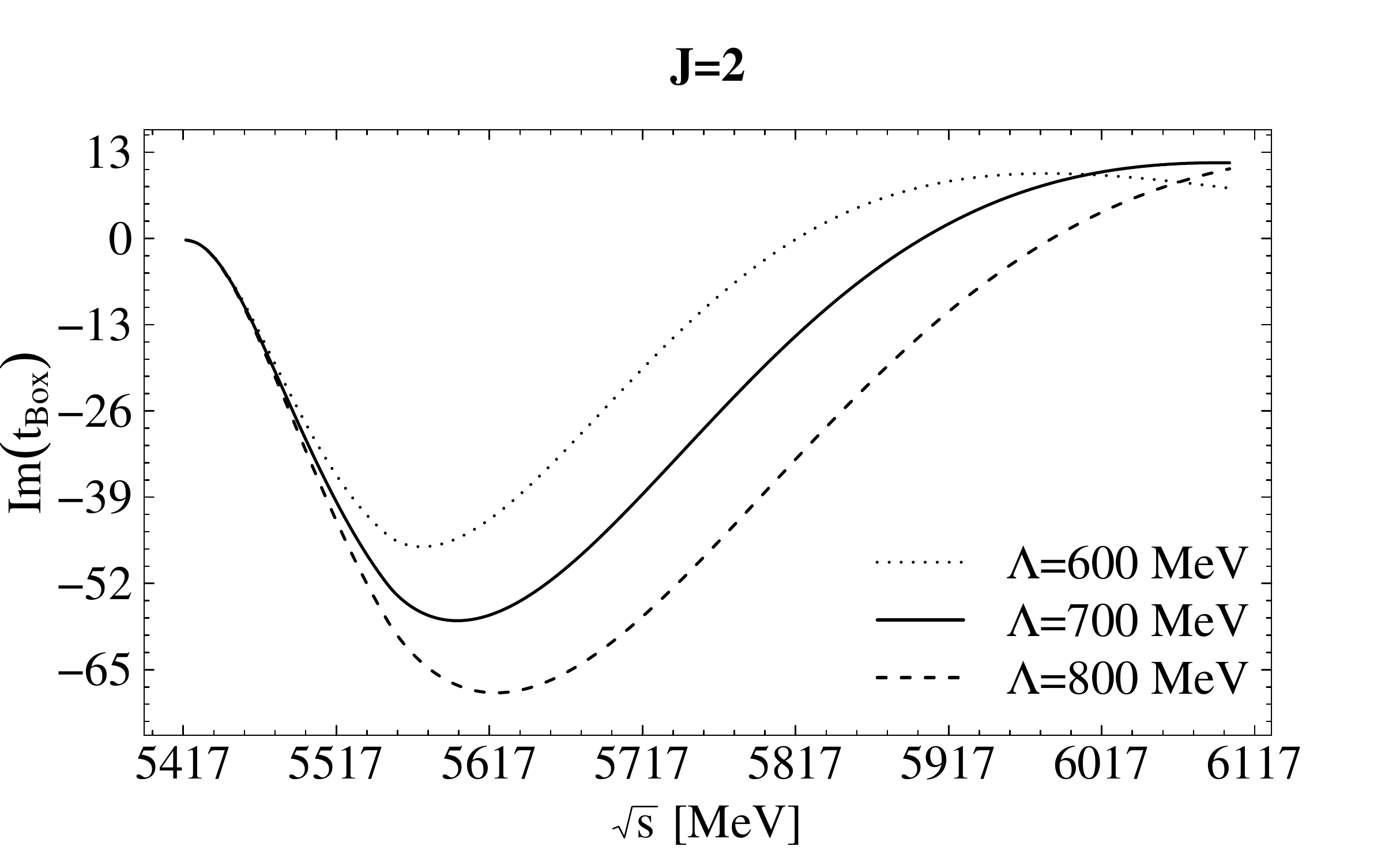}
  \caption{The imaginary part of box potential for $(I,J)=(1/2,0)$ and $(I,J)=(1/2,2)$.\label{imaginarypartbox}}
\end{figure*}
where
\begin{align}
f(P^0,\vec{q}^{\,2})&=4\left\lbrace-32k^0_3P^0\omega_\pi^2\omega_B((P^0)^2-2\omega_\pi^2-3\omega_\pi\omega_B-\omega_B^2)\right.\nonumber\\
&+2(k_3^0)^3P^0\omega_B((P^0)^2-5\omega_\pi^2-2\omega_\pi\omega_B-\omega_B^2)\nonumber\\
&+(k_3^0)^4(2\omega_\pi^3-(P^0)^2\omega_B+3\omega_\pi^2\omega_B+2\omega_\pi\omega_B^2+\omega_B^3)\nonumber\\
&+4\omega_\pi^2(8\omega_\pi^5+33\omega_\pi^4\omega_B+54\omega^3\omega_B^2+3\omega_B((P^0)^2\nonumber\\
&-\omega_B^2)^2+18\omega_\pi\omega_B^2(-(P^0)^2+\omega_B^2)+\omega_\pi^2(-12(P^0)^2\omega_B\nonumber\\
&+44\omega_B^3))-(k_3^0)^2(16\omega^5+63\omega^4\omega_B+74\omega_\pi^3\omega_B^2\nonumber\\
&+\omega_B((P^0)^2-\omega_B^2)^2+32\omega_\pi^2\omega_B(-(P^0)^2+\omega_B^2)\nonumber\\
&\left.+\omega_\pi(-6(P^0)^2\omega_B^2+6\omega_B^4))\right\rbrace,
\end{align}
$\omega_\pi = \displaystyle\sqrt{\vec{q}^{\,2}+m_\pi^2}$, $\omega_B = \displaystyle\sqrt{\vec{q}^{\,2}+m_B^2}$, and $P^0=k_1^0+k^0_2$.
\begin{figure*}
  \includegraphics[scale=0.35]{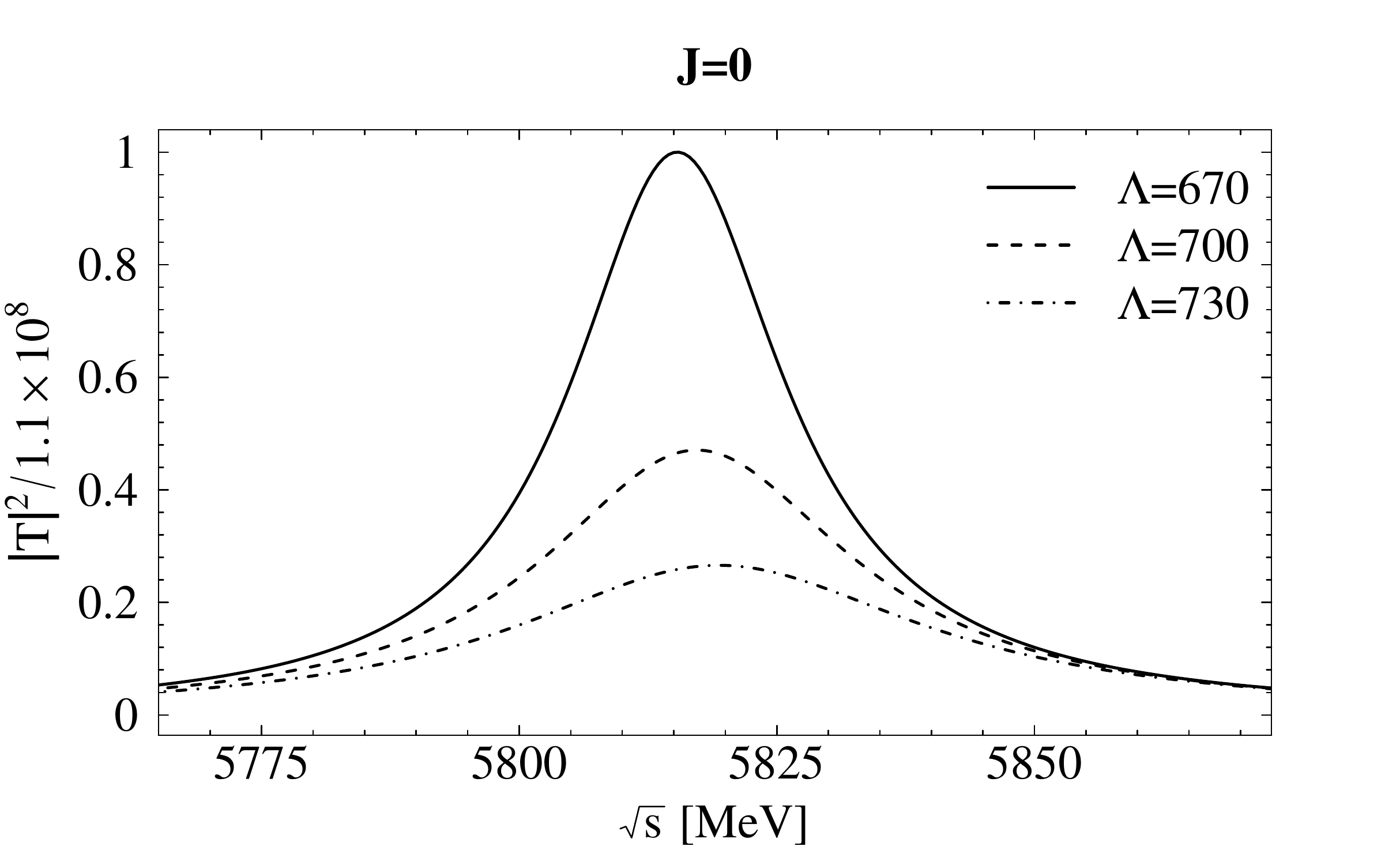}
  \includegraphics[scale=0.35]{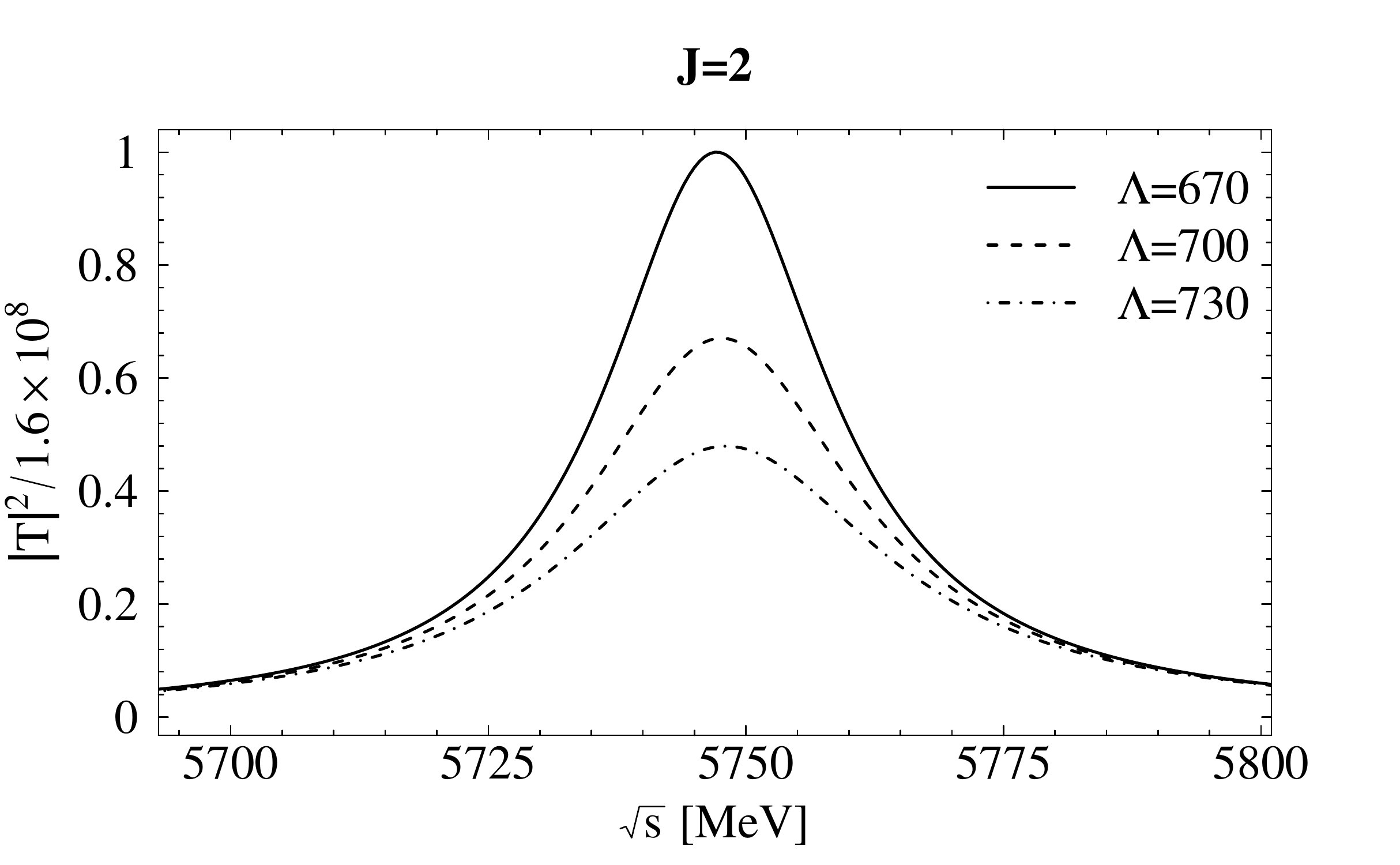}
\caption{Squared amplitudes for $J=0$ and $J=2$ which depend on the energy in the center of mass including the convolution of $\rho$ mass distribution and the box diagram.}\label{Tsquarebox}
\end{figure*}
Here, in order to calculate the box diagram amplitude, one has first in\-te\-gra\-ted analytically the $q^0$ variable. Note that the integral is logarithmically divergent, and as in \cite{hidekoraquel} we use a form factor to regularize the loop in addition to the $q_{max}$ value used before.
The spin structure only allows $J=0$ and $2$. The reason why $J=1$ is forbidden is that the parity of $\rho B^*$ system is positive with s wave, and the angular momentum of $\pi B$ system has to be $L=0, 2$. Therefore, the spin of $\pi B$ would be 0 or 2, but not 1. Using again the results of \cite{hidekoraquel} we find the spin projections
\begin{align}
\delta V^{\pi B,I=1/2,J=0}=20\tilde{V}^{(\pi B)}, \quad \delta V^{\pi B,I=1/2,J=2}=8\tilde{V}^{(\pi B)},
\end{align}
where $\tilde{V}^{(\pi B)}$ is given in Eq. (\ref{boxintegral}) after removing the polarization vectors. In this work, we also use a form factor in each vertex of the box diagram, and then finally, $g^4$ is replaced with
\begin{align}
g^2_{\rho \pi\pi}g_{B^*B\pi}^2(e^{-\vec{q}^{\,2}/\Lambda^2})^4
\end{align}
where $g_{\rho\pi\pi}\equiv g=m_{\rho}/(2f_\pi)$ and $g_{B^*B\pi}=gm_{B^*}/m_{K^*}$ (see Eq. (\ref{eq:g})), and $\Lambda$ is of the order of 1 GeV.

The real part of the box diagram contribution is neglected, since it is very small compared with those of the contact and vector exchange terms as we can see in Fig. \ref{realpartbox}. The imaginary part that we focus on is shown in Fig. \ref{imaginarypartbox}. If $\Lambda$ is taken as $0.67$ GeV and $q_{max}$ as 1.3 GeV, the width for $J=2$ is 25.5 MeV which is in agreement with the experimental value in the PDG. For $J=0$ the width is then 24.7 MeV, while the state with $J=1$ has no width in our approach. If $\Lambda$ is increased to $0.73$ GeV, we obtain a width for $J=2$ of 37.5 MeV, and 47.8 MeV for $J=0$.
We see that we can obtain a width comparable to experiment using cut offs or form factors of natural size.

In Fig. \ref{Tsquarebox} we show the line shape of $\vert T\vert^2$ including the box diagram, which should be compared to Fig. \ref{TsquareRhoConvolution}. We can see that for $J=0, 2$ a width for the states appears but the peak does not move. On the other hand, for $J=1$ we still have the results of Fig. \ref{TsquareRhoConvolution}, since as discussed above, in this case there is no box diagram.

\subsection{$\rho B$ system}
As we have mentioned in the previous subsection, in the PDG the mass of $B_1(5721)$ is smaller than that of $B^*_2(5747)$. However, for the $\rho B^*$ systems the mass of the $J=1$ state is larger than that of the $J=2$ state. Henceforth, we turn to the $\rho B$ system and investigate its interaction.

\begin{figure}[htb]
    \includegraphics[scale=0.35]{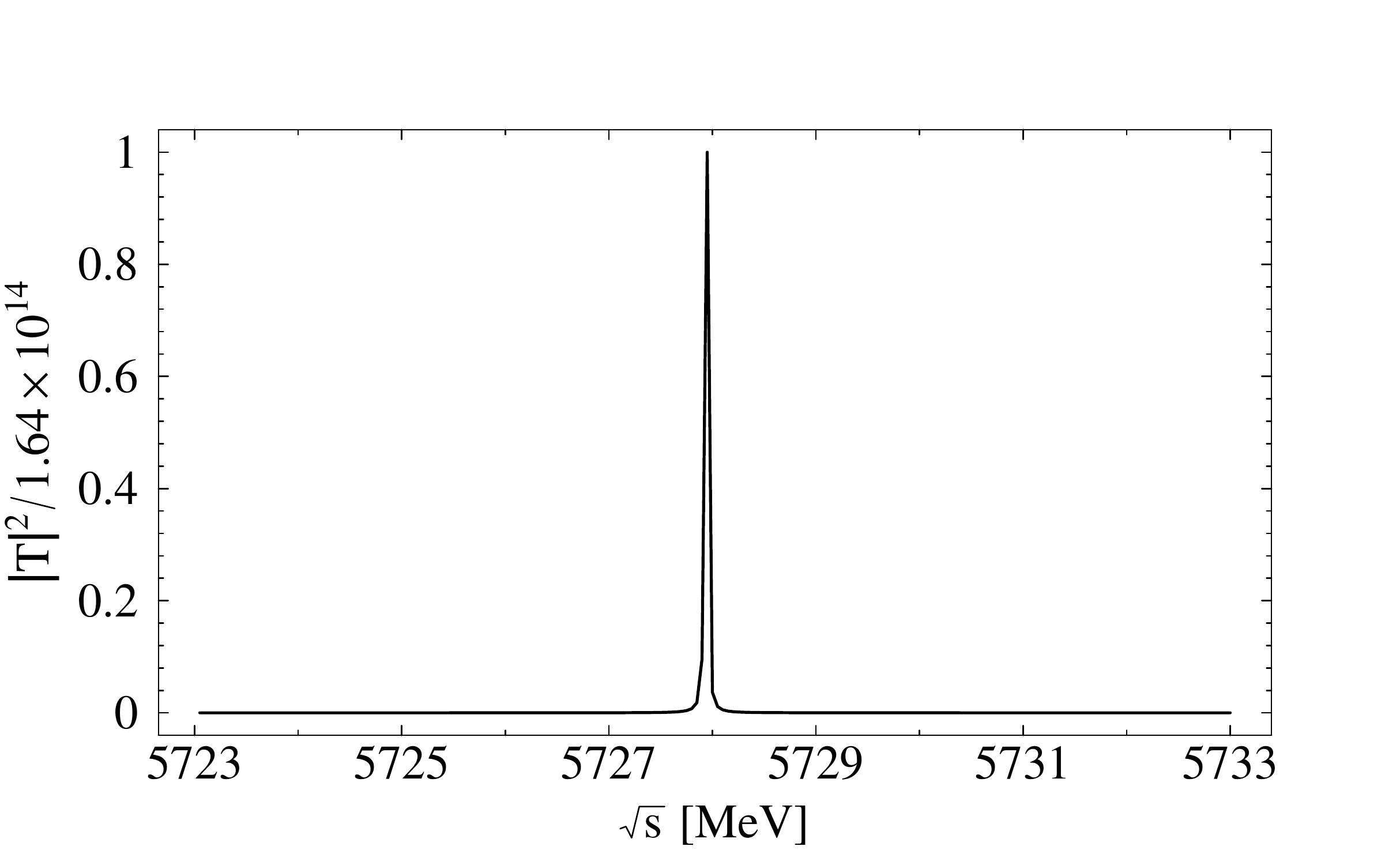}
\caption{Squared amplitude for $\rho B$ sector with spin 1.\label{rhoBTmatrix}}
\end{figure}
For this system there are no contact terms, but we have the vector exchange terms only. In addition, the $\omega$ channel is now inoperative since the $\rho \rho \omega$ vertex is zero by G-parity and $\omega \omega \rho$ is zero by C-parity and isospin. Note that in the case of the vector-vector interaction it is the exchange term of Fig. \ref{contact} (c) the one that makes $\omega B^*$ mix with $\rho B^*$. The equivalent diagrams would involve anomalous terms which are small. In any case the factor $m_V^2/m_{B^*}^2$ of these terms renders them negligible, of the order of $1\%$ also in the case of the vector-vector interaction.

\begin{figure}[htb]
	\centering
	{\includegraphics[scale=0.7]{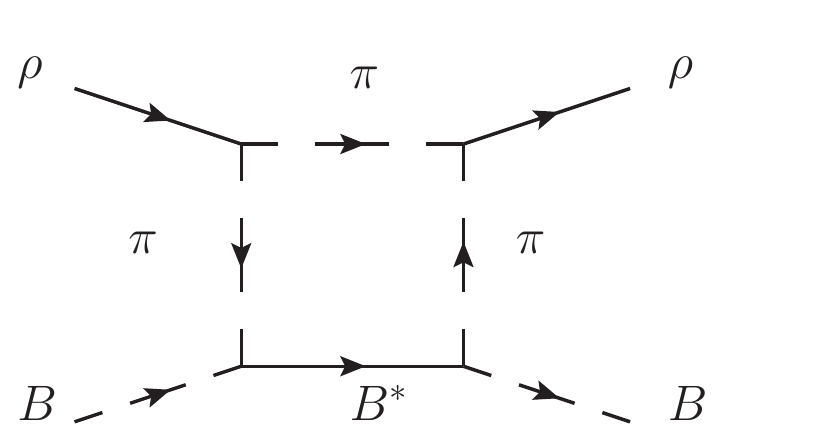}}
\caption{Box diagram for $\rho B\to \rho B$ with $B^*\pi$ intermediate state.}
\label{fig:box-4}	
\end{figure}

Since the strength of the interaction is the same as in the $\rho B^*\to \rho B^*$ case we expect to find a bound state as before. If the cut off $q_{max}$ in the G function is taken as 1.3 GeV, we find the pole position at $5728$ MeV (see Fig. \ref{rhoBTmatrix}), which is consistent with the PDG value of $B_1(5721)$. The coupling to $\rho B$ channel is also computed, and found to be $g_{\rho B}=41.6$ GeV. It is very interesting to also calculate the width of this state. The PDG does not quote any number but it states that the dominant decay mode is $B^*\pi$. This comes out naturally in our approach by means of the box diagram of Fig. \ref{fig:box-4} .

It is easy to see the contribution for this new box diagram following the steps of \cite{hidekoraquel}. There we had the combination

\begin{align}
&\int d^3q \epsilon_i^{(1)}\epsilon_j^{(2)}\epsilon_l^{(3)}\epsilon_m^{(4)}q_iq_jq_lq_m \mathcal{F}(\vec{q}^{\,2})\nonumber\\
=&\frac{1}{15}\int d^3q \epsilon_i^{(1)}\epsilon_j^{(2)}\epsilon_l^{(3)}\epsilon_m^{(4)}\vec{q}^4(\delta_{ij}\delta_{em}\nonumber\\
&+\delta_{ie}\delta_{jm}+\delta_{im}\delta_{je}) \mathcal{F}(\vec{q}^{\,2})\nonumber\\
=&\frac{1}{15}\left(5\mathcal{P}(0)+2\mathcal{P}(2)\right)\int d^3q \vec{q}^{\,4}\mathcal{F}(\vec{q}^{\,2}) ,
\label{Eq:structure-box-D-rho-Bx}
\end{align}
here $\mathcal{F}(\vec{q}^{\,2})$ is a function depending on the square of the three momentum $\vec{q}^{\,2}$, the center of mass energy and the masses of the mesons appearing in Fig. \ref{fig:box-4} .

Now we have the same original form as in the beginning of the equation but we must sum over the $B^*$ polarization of the intermediate $B^*$ state. Since we are only concerned about the imaginary part, the on shell approximation for the intermediate $B^*$ is sufficient and the contraction of $\epsilon_j^{(2)}\epsilon_m^{(4)}$ gives $\delta_{ij}$. Then the remaining structure is 
\begin{align}
&\int d^3q \epsilon_i^{(1)}\epsilon_l^{(3)}q_iq_l\vec{q}^{\,2}\displaystyle{\tilde{\mathcal{F}}}(\vec{q}^{\,2})
=\epsilon_i^{(1)}\epsilon_l^{(3)}\int d^3q \frac{1}{3}\vec{q}^{\,4}\displaystyle{\tilde{\mathcal{F}}}(\vec{q}^{\,2}) ,
\end{align}
where $\displaystyle\tilde{\mathcal{F}}(\vec{q}^{\,2})$ has the same form as $\mathcal{F}(\vec{q}^{\,2})$ after making the change $m_{B^*}\to m_{B}$ and $m_{B}\to m_{B^*}$, up to a constant factor that we shall discuss right now. The interaction Lagrangian of Eq. (\ref{lagrangian2}) involves derivatives of the pseudoscalar fields. In comparison with the previous situation which is depicted in the box diagram of Fig. \ref{box}, now the $B^*B\pi$ vertex does not have a $B$ meson carrying the $q$ momenta of the integral, since this meson is external (see Fig. \ref{fig:box-4}). Before we had in the incoming $B^*B\pi$ vertex a factor
\begin{align}
  \propto \left((P-q)+(k_1-q)\right)_\mu  \xrightarrow{\vec{k}_i\rightarrow 0}  -2q_i
\end{align}
corresponding to the momentum of the $B$ and $\pi$ internal mesons in Fig. \ref{box}. Now in Fig. \ref{fig:box-4} the incoming $B^*B\pi$ vertex is 
\begin{align}
\propto \left(-k_2+(k_1-q)\right)_\mu \xrightarrow{\vec{k}_i\rightarrow 0}   -q_i
\end{align}
because the derivatives involve the external pseudoscalar $B$ and the internal $\pi$. As a consequence the amplitudes will lack a factor two in each of the $B^*B\pi$ vertex, so $\tilde{\mathcal{F}}=\mathcal{F}/4$. When we remove the polarization vectors the structure is 
\begin{align}
\int d^3q \frac{1}{3}\vec{q}^{\,4}\displaystyle{\tilde{\mathcal{F}}}(\vec{q}^{\,2})=\int d^3q \frac{1}{3}\vec{q}^{\,4}{\mathcal{F}}(\vec{q}^{\,2})\frac{1}{4}.
\end{align}
Hence, comparing with Eq. (\ref{Eq:structure-box-D-rho-Bx}) we see that the strength of the $\rho B$ box potential is identical to the former one with $J=0$ of the $\rho B^*$,  divided by four (changing the intermediate mass of the $B$ to the present one of $B^*$ and vice versa). 
\begin{figure}[htb!]
	\centering
	\includegraphics[scale=0.35]{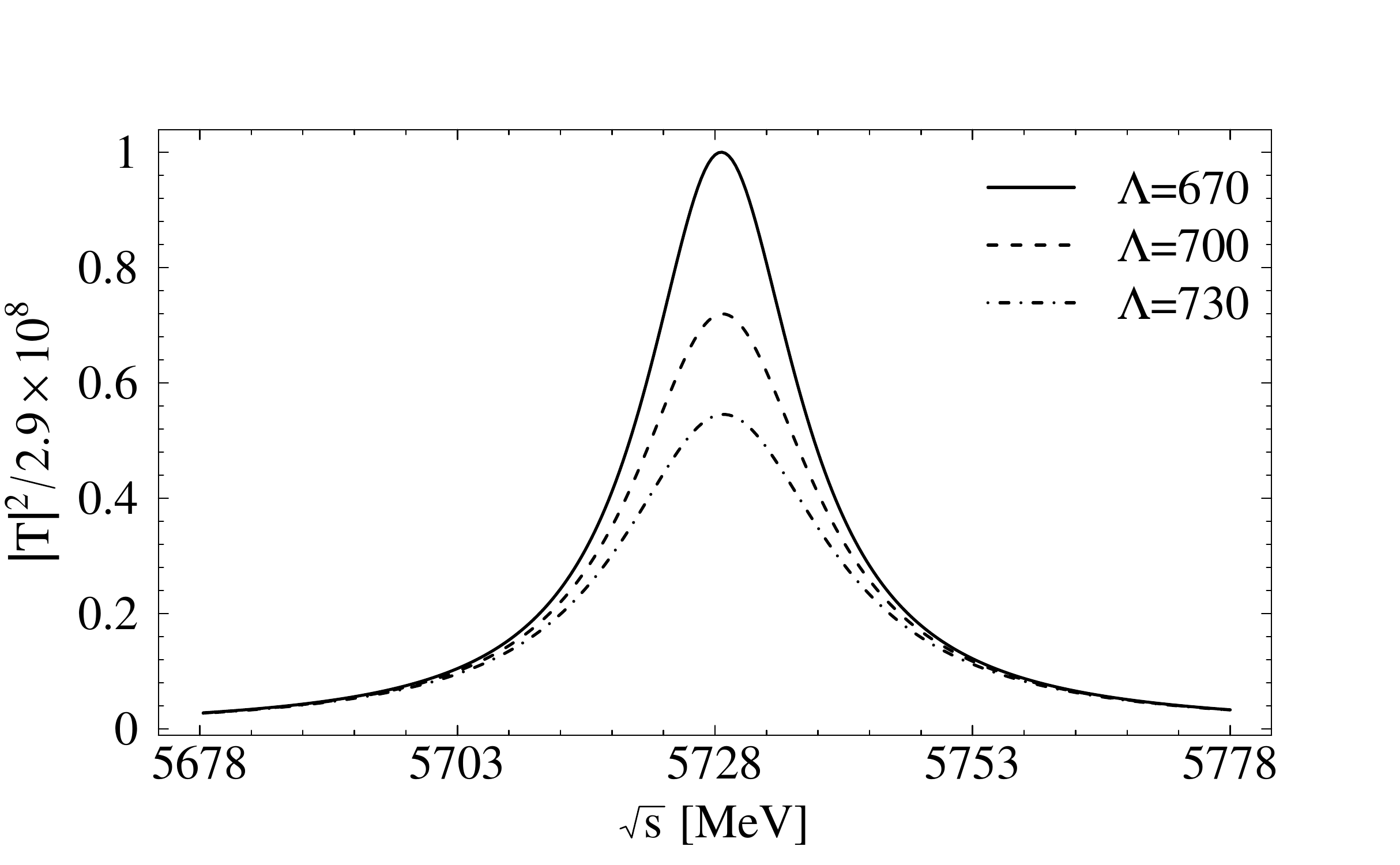}
\caption{Squared amplitude of $\rho B$ system which depends on the c.o.m. energy including the convolution of the $\rho$ mass distribution and the box diagram contribution.}
\label{fig:rhoBbox}
\end{figure}

In Fig. \ref{fig:rhoBbox} we plot $\vert T\vert^2$ for this case with the same parameters used before to obtain the width of the $B_2^*(5747)$. We see that we obtain a width around 20 MeV, which is a prediction of the present work.

\section{Summary}
In this work we have studied the $\rho B^*$, $\omega B^*$ and $\rho B$ interactions by using the local hidden gauge unitary approach. First we have solved the Bethe-Salpeter equation in coupled channels for the $\rho B^*$ and the $\omega B^*$ sectors, using the tree level amplitudes and re\-gu\-la\-ri\-zing the loop function with a cut off of $1.3$ GeV. In this way we have found three bound states, with masses 5812, 5817 and 5745 MeV for $I=1/2$ and $J=0, 1, 2$, respectively, identifying the $J=2$ state with the $B_2^*(5747)$ \cite{Agashe:2014kda} of mass $5743\pm 5$ MeV. Despite having considered the rho mass distribution, all the states that we have found show small widths.
In order to generate the correct width of the state with $J=2$ as that of the experimental $B_2^*(5747)$, which is quoted as $23^{+5}_{-11}$ MeV, we have taken into account the box diagram mediated by the $\pi B$ which accounts for this decay channel.
We have also considered a form factor for the off-shell pions and a rescaled coupling in the $B^* B\pi$ vertex. In this way, we have obtained the widths $25.5\sim 37.5$ MeV for $J=2$ and $24.7\sim 47.8$ MeV for $J=0$, taking $\Lambda=0.67\sim 0.73$ GeV, $q_{max}=1.3$ GeV. Since the pole position of $J=1$ is larger than that of $J=2$, while in the PDG there is a spin one state $B_1(5721)$ which mass is smaller than the $B_2^*(5747)$ mass, we have considered the $\rho B$ system.

For the $\rho B$ interaction in the local hidden gauge approach we have found a bound state of mass $5728$ MeV, which is consistent with the experimental value of the $B_1(5721)$. We have also predicted a width for this state considering the box diagram contribution in a similar manner as for the $\rho B^*$ system. The width that we have obtained is around $20$ MeV. We su\-mma\-ri\-ze our results in Table \ref{Table:summary}.
\begin{table}[hbt!]
\centering
\caption{Summary of the states found in the $\rho(\omega) B^*$ and $\rho B$ interaction.\label{Table:summary}}
	\begin{tabular}{cccccc}
		\toprule[1pt]		
Main    & $I(J^P)$ & $M$ [Mev] & $\Gamma$ [MeV] & Main decay & Exp $(M,\Gamma)$\\
channel &		 &	     &                & channel    & [MeV]\\
 		\hline
$\rho B^*$ & $\frac{1}{2}(0^+)$ & $5812$ & $25-45$ & $\pi B$      & \\ 
$\rho B^*$ & $\frac{1}{2}(1^+)$ &  $5817$ & $0$&  $ $             &\\ 
$\rho B^*$ & $\frac{1}{2}(2^+)$ & $5745$ & $25-35$ & $\pi B$      & $(5743\pm5\,,\,23^{+5}_{-11})$\\ 
$\rho B$ & $\frac{1}{2}(1^+)$ & $5728$ & $18-24$ & $\pi B^*$      & $(5723.5\pm 2\,,\,-)$\\ 
	\bottomrule[1pt]
	\end{tabular}	
\end{table}

Finally we have investigated if there is some aspect in the interaction which can be related to the heaviness of the system under consideration. The fact that the $B$ mesons have a large mass can justify the study of the $\rho B$ and $\rho B^*$ systems under the frame of heavy quark spin symmetry. We have splitted these states in terms of eigenstates of total angular momentum of the light quarks as in \cite{Xiao:2013yca}. 

We find that the dominant terms in our approach, due to light vector exchange, which go like $\mathcal{O}\left(\frac{1}{m_{B^*}}\right)^0$ fulfill the LO constrains of HQSS, while the contact terms and those coming from the exchange of $B^*$ are subdominant ($\mathcal{O}\left(\frac{1}{m_{B^*}}\right)^0$) and do not fulfill the LO HQSS rules. While in the $D\rho$ sector these terms were not too small, in the present case they are much smaller and we have a near degeneracy in the $\rho B^*$ states with $J=0,1,2$.

\section*{Acknowledgments}
This work is partly supported
by the Spanish Ministerio de Economia y Competitividad and European
FEDER funds under the contract number FIS2011-28853-C02-01,
FIS2011-28853-C02-02, FIS2014-57026-REDT, FIS2014-51948-C2-1-P, and FIS2014-51948-C2-2-P, and the Generalitat Valenciana in the program
Prometeo II-2014/068. We acknowledge the support of the European
Community-Research Infrastructure Integrating Activity Study of
Strongly Interacting Matter (acronym HadronPhysics3, Grant Agreement
n. 283286) under the Seventh Framework Programme of EU.

\bibliographystyle{plain}

\end{document}